\documentclass[11pt,a4paper]{article}
\usepackage[utf8]{inputenc}
\usepackage[expansion=false]{microtype}
\usepackage{amsmath,amssymb,graphicx,booktabs}
\usepackage[margin=23mm,top=21mm,bottom=23mm]{geometry}
\usepackage{hyperref}
\hypersetup{colorlinks=true,linkcolor=black,citecolor=black,urlcolor=black,
pdftitle={Horizon Response to Orbital Redistribution in Self Consistent Einstein--Vlasov Black Hole Environments},
pdfauthor={Anirudh Pradhan, K. Ghaderi, M. Zeyauddin, Ajit Kumar}}
\begin{document}
\begin{center}
{\Large Horizon Response to Orbital Redistribution in Self Consistent Einstein--Vlasov Black Hole Environments}\\[0.8em]
{\normalsize Anirudh Pradhan$^{1}$, K. Ghaderi$^{2,*}$, M. Zeyauddin$^{3}$, and Ajit Kumar$^{4}$}\\[0.45em]
{\small $^{1}$Centre for Cosmology, Astrophysics and Space Science (CCASS), GLA University, Mathura 281406, U.P., India}\\
{\small $^{2}$Department of Physics, Mari.C., Islamic Azad University, Marivan, Iran}\\
{\small $^{3}$Department of General Studies (Mathematics), Jubail Industrial College, Jubail 31961, Saudi Arabia}\\
{\small $^{4}$Department of Mathematics, Faculty of Engineering, Teerthanker Mahaveer University, Moradabad-244001, India}\\[0.35em]
{\small $^{*}$Corresponding author: \texttt{k.ghaderi@iau.ac.ir}}
\end{center}

\begin{abstract}
We study whether orbital redistribution in a self consistent spherical Einstein--Vlasov environment can change the asymptotic normalization of a black hole horizon while the horizon mass, particle rest mass and ADM mass are all held fixed. The matter variable is the occupation of regular bound orbital actions, so the stress tensor, orbital energies and gravitational field are determined by the same distribution. Linearization of the spherical mass constraint gives a kinetic variation law in which orbital energy is conjugate to occupation and the inner lapse ratio multiplies the horizon mass variation. On a locally regular equilibrium branch, mixed variations imply an integrated reciprocity relation between the central mass derivative of orbital energy and the directional response of the horizon normalization. We construct finite mass populations and trace constant mass redistribution curves. For the reference rest mass ratio $0.30$, the calculated endpoint surface gravity shifts are approximately $+15.37$ and $-15.33$ parts per million. The constrained response cancels at simultaneous dilute and Newtonian order, is nonzero for exact Schwarzschild orbits, and is further modified by self gravity. Independent balance, integral, derivative and refinement tests resolve the response well below its reported magnitude.
\end{abstract}

\section{Introduction}
\label{sec:introduction}

The surface gravity of a black hole surrounded by matter depends not only on the local horizon geometry but also on the normalization of the stationary Killing field relative to infinity. In spherical symmetry, a vacuum interval adjacent to the horizon fixes the local Schwarzschild geometry through the horizon mass, whereas the relation between the local Killing time and an asymptotic clock depends on the intervening matter. The mechanics of stationary black holes with exterior matter has a long history \cite{Bardeen1973}. The Noether charge formulation provides a general setting for the first law \cite{IyerWald1994}, while explicit matter variations require a specification of the variables and constraints used to compare neighbouring configurations \cite{Iyer1997}. For collisionless matter, that specification must include how particles are distributed among bound orbits.

Environmental changes of horizon normalization are already present in spherical black hole thermodynamics \cite{Visser1992}. In particular, a thin shell can change the surface gravity while the horizon and total gravitational masses are prescribed. Related environmental redshifts also enter black hole observables in matter distributions \cite{Cardoso2022}. The question considered here is more restrictive: can a redistribution of a collisionless population change the horizon normalization when the horizon mass, particle rest mass and ADM mass are all fixed, with the stress tensor and metric determined self consistently from the same phase space distribution?

The Einstein--Vlasov system is a natural setting for this problem because the energy density and principal pressures are moments of one distribution function rather than independent input profiles. The mathematical and physical structure of the system is reviewed in Ref.~\cite{Andreasson2011}. Smooth static spherical equilibria were established by Rein and Rendall \cite{ReinRendall1993}, and static solutions with finite matter support, including configurations with a central Schwarzschild singularity, were constructed by Rein \cite{Rein1994}. More recently, Jabiri constructed static spherical Einstein--Vlasov families bifurcating from Schwarzschild with compact matter support separated from the horizon \cite{Jabiri2021}. For collisionless matter on a prescribed Schwarzschild background, bound orbit constructions provide explicit particle currents, stress tensors and global integrals \cite{Gabarrete2023}. Action angle variables have also recently been used to formulate a linear stability criterion for Einstein--Vlasov matter shells surrounding a Schwarzschild black hole \cite{Gunther2025}; the present use is instead to define admissible occupation variations and the associated horizon response. These results supply the equilibrium and kinetic background; a finite mass response calculation must additionally include the field generated by the population itself.

Orbital actions provide a direct way to state which microscopic information is preserved in a comparison of nearby equilibria. Adiabatic growth of a central black hole has been used to study dark matter spikes \cite{Gondolo1999}, including relativistic distributions obtained by preserving actions \cite{Sadeghian2013}. Action based distribution functions are also standard in stellar dynamics \cite{Posti2015}. In relativistic two body mechanics, actions likewise enter first laws through their conjugate frequencies \cite{LeTiec2015}. In the present spherical setting, the radial action and total angular momentum label the occupied bound tori after phase and orientation averaging. Holding their dimensional occupation fixed describes adiabatic transport of the same population. Holding actions scaled by a changing central mass fixed instead defines a different family.

We use the action occupation as the matter variable on a locally regular branch of static equilibria. Its integral is the total particle rest mass, and its compact support is kept away from capture and separatrix boundaries. The canonical variational formulation of collisionless gravitating matter provides the appropriate framework \cite{Andersson2019}. Linearizing the spherical mass constraint gives an explicit kinetic variation law: the horizon contribution is controlled by the inner lapse ratio, while an occupation perturbation is weighted by the orbital energy measured at infinity. For smooth occupation perturbations, commutation of the horizon mass and occupation variations gives an integrated reciprocity relation between the mass derivative of orbital energy and the directional response of the horizon normalization. The numerical tests below use precisely these projected identities, avoiding any assumption of a pointwise response kernel beyond the regularity required by the computed finite dimensional occupation family.

This source based formulation is important when discussing black hole environments. A prescribed density profile alone does not determine a relativistic matter source or its admissible variations, as emphasized in recent consistency analyses of halo metrics \cite{Bolokhov2026}. Thermodynamic calculations in prescribed rotating halo geometries also illustrate the dependence of horizon quantities on the chosen matter completion \cite{Araujo2026}. Here the local momentum integrals determine the density, radial pressure and tangential pressure, while the same metric determines the orbital energies entering those integrals. The resulting equilibria therefore retain both the anisotropic kinetic stresses and the gravitational response of the population.

The central application is a constrained redistribution within a compact action population. Three occupation directions vary the mean radial action, the mean angular momentum and the radial action variance while preserving the particle rest mass. Orbital energy integrals give the corresponding ADM mass gradients. Two directions are then combined to construct a curve along which the horizon mass, particle rest mass and ADM mass are all fixed. A nonzero constrained derivative of the inner lapse ratio changes the surface gravity at fixed horizon area. For the reference population with rest mass $0.30M_\bullet$, the calculated endpoint shifts are approximately $+15.37$ and $-15.33$ parts per million relative to the undeformed equilibrium. Every point on this curve is reconstructed as a separate solution of the coupled equilibrium problem.

An independent dilute calculation separates the effect from its leading Newtonian counterpart. In the Kepler limit, the dependence of orbital energy on $J_r+L$ makes the constrained horizon response cancel at simultaneous dilute and Newtonian order. Exact Schwarzschild orbits break this proportionality and produce a nonzero dilute response. The first relativistic correction identifies the leading orbital scale dependence, while the coupled equilibria quantify the additional modification produced by self gravity over several action distributions. Local stress balance, conserved mass integrals, action reconstruction, independent orbital energy integrals, finite differences and Hessian symmetry provide mutually distinct checks of the result.

Section~\ref{sec:kinetic} derives the kinetic mass variation law and the orbital benchmark. Section~\ref{sec:selfgravity} constructs the coupled equilibrium families and tests their nonlinear response. Section~\ref{sec:redistribution} develops occupation redistribution under the three mass constraints and compares the horizon response with its dilute and weak field limits. Section~\ref{sec:discussion} discusses the physical interpretation, and Section~\ref{sec:conclusions} summarizes the main results.

\section{Collisionless equilibria and constrained horizon variations}
\label{sec:kinetic}

The response of a black hole surrounded by collisionless matter depends on which properties of the particle population are preserved. We formulate this dependence using a rest mass distribution on canonical phase space. The metric, the orbital Hamiltonian and the matter moments are then determined by the same distribution. This is the setting of the Einstein--Vlasov system \cite{Andreasson2011} and its canonical variational formulation \cite{Andersson2019}. Smooth, finite matter shells separated from a Schwarzschild horizon provide an established class of nearby equilibria \cite{Jabiri2021}.

A prescribed density and a relativistic matter model carry different information. In particular, recent source consistency analyses show why a halo profile does not by itself determine its relativistic completion \cite{Bolokhov2026}. Thermodynamic calculations for prescribed rotating halo geometries \cite{Araujo2026} consequently motivate a specification of the admissible matter variations. Here the population is labelled by orbital actions. Adiabatic action preservation has already been used in relativistic dark matter spike calculations \cite{Sadeghian2013}; our formulation retains the gravitational field of the population in the equilibrium and variation equations.

\subsection{Geometry and canonical matter variables}
\label{sec:geometry}

We use $G=c=1$ and the metric signature $(-,+,+,+)$. A static, spherical exterior is written as
\begin{equation}
 ds^2=-\alpha(r)^2dt^2+B(r)^{-1}dr^2+r^2d\Omega^2,
 \qquad B(r)=1-\frac{2m(r)}r,
 \label{eq:metric}
\end{equation}
with $\alpha(\infty)=1$ and $m(\infty)=M_\infty$. The horizon mass is $M_\bullet=r_h/2$. The matter has compact support in $R_{\rm in}<r<R_{\rm out}$, where $R_{\rm in}>r_h$, and $B>0$ throughout the matter region. Its occupied trajectories lie on the outer bound orbital component of the timelike phase space. This component restriction excludes trajectories with the same energy and angular momentum that instead reach the horizon \cite{Jabiri2021}.

Let $u_i$ be the covariant spatial momentum per unit particle rest mass, and let $q_{ij}$ be the spatial metric induced by Eq.~\eqref{eq:metric}. We absorb the fixed particle rest mass into the distribution $f$, so that
\begin{equation}
 d\Gamma=d^3x\,d^3u,\qquad
 M_{\rm rest}=\int f\,d\Gamma,\qquad
 \varepsilon=\sqrt{1+q^{ij}u_i u_j},\qquad H=\alpha\varepsilon.
 \label{eq:canonical}
\end{equation}
Here $\varepsilon$ is the local specific energy and $H=-u_t$ is the specific energy measured at infinity. The phase space coordinates are canonical after the constant rest mass normalization. With $q=\det(q_{ij})$, the local energy density and principal pressures are
\begin{align}
 \rho&=\frac1{\sqrt q}\int f\varepsilon\,d^3u,\qquad
 p_r=\frac1{\sqrt q}\int f\frac{B u_r^2}{\varepsilon}\,d^3u,
 \label{eq:moments1}\\
 p_t&=\frac1{2\sqrt q}\int f\frac{L^2}{r^2\varepsilon}\,d^3u,
 \qquad L^2=u_\theta^2+\frac{u_\phi^2}{\sin^2\theta}.
 \label{eq:moments2}
\end{align}
Consequently, $p_r,p_t\geq0$ and
\begin{equation}
 \rho-p_r-2p_t=\frac1{\sqrt q}\int\frac f\varepsilon\,d^3u\geq0.
 \label{eq:trace}
\end{equation}
No independent pressure closure is introduced. The kinetic and Einstein equations are
\begin{align}
 \partial_t f+\{f,H\}&=0,\qquad
 \{a,b\}=\frac{\partial a}{\partial x^i}\frac{\partial b}{\partial u_i}
 -\frac{\partial a}{\partial u_i}\frac{\partial b}{\partial x^i},
 \label{eq:vlasov}\\
 m'&=4\pi r^2\rho,\qquad
 \frac{\alpha'}\alpha=\frac{m+4\pi r^3p_r}{r^2B}.
 \label{eq:einstein}
\end{align}
Stationarity gives $\{f,H\}=0$. The remaining radial conservation equation follows from these equations and the kinetic moments.

Define the lapse ratio $Z=\alpha/\sqrt B$. Equation~\eqref{eq:einstein} gives
\begin{equation}
 \frac{Z'}Z=\frac{4\pi r(\rho+p_r)}B,\qquad
 Z(r)=\exp\!\left[-\int_r^\infty
 \frac{4\pi s[\rho(s)+p_r(s)]}{B(s)}\,ds\right].
 \label{eq:Z}
\end{equation}
Inside the vacuum gap, $m=M_\bullet$ and $Z=Z_h$ is constant. Thus
\begin{equation}
 \alpha=Z_h\sqrt{1-2M_\bullet/r},\qquad
 \kappa_h=\frac{Z_h}{4M_\bullet},\qquad
 A_h=16\pi M_\bullet^2.
 \label{eq:horizon}
\end{equation}
The normalization at infinity is essential: $Z_h$ is a physical redshift relative to that normalization. Environmental redshift is an established effect in black hole halo geometries \cite{Cardoso2022}. Here Eq.~\eqref{eq:Z} also identifies the radial pressure contribution for an eccentric collisionless population. Positivity of the moments implies $0<Z_h\leq1$ for the regular configurations under consideration.

\subsection{Orbit occupations and admissible variations}
\label{sec:actions}

On a regular bound torus, choose specific actions
\begin{equation}
 J_r=\frac1\pi\int_{r_p}^{r_a}\frac{dr}{\sqrt B}
 \sqrt{\frac{E^2}{\alpha^2}-1-\frac{L^2}{r^2}},\qquad
 J_\theta=L-|L_z|,\qquad J_\phi=L_z,
 \label{eq:actions}
\end{equation}
where $E=H$ and $r_p,r_a$ are the turning points in the occupied component. A spherical, phase mixed distribution has $f=\mathcal F(J_r,L)$ and is independent of the orientation of the orbital plane. The action support is compact and remains separated from capture and separatrix boundaries. All dimensional actions and the particle rest mass are held fixed when comparing the same population. Keeping $J_r/M_\bullet$ and $L/M_\bullet$ fixed would define a different variation.

Integrating over the angle variables and orbital orientations gives
\begin{equation}
 M_{\rm rest}=\int n(J_r,L)\,dJ_r\,dL,\qquad
 n(J_r,L)=2L(2\pi)^3\mathcal F(J_r,L).
 \label{eq:occupation}
\end{equation}
The factor $2L$ is the integral over $-L<L_z<L$. The same population can be used to reconstruct the local moments. Define
\begin{equation}
 D(r)=E^2-\alpha^2(1+L^2/r^2),\quad
 T_r=2\int_{r_p}^{r_a}\frac{E\,dr}{\alpha\sqrt B\sqrt D},\quad
 P(r|J_r,L)=\frac{2E}{T_r\alpha\sqrt B\sqrt D},
 \label{eq:probability}
\end{equation}
where $P=0$ outside the radial interval and $\int P\,dr=1$. This is the probability density in radius for sampling uniformly in the asymptotic time $t$. Then
\begin{align}
 4\pi r^2\rho(r)&=\frac1{Z(r)}\int n E P\,dJ_r\,dL,
 \label{eq:orbitalrho}\\
 4\pi r^2p_r(r)&=\frac1{Z(r)}\int n E P\frac{D}{E^2}\,dJ_r\,dL.
 \label{eq:orbitalpr}
\end{align}
The tangential pressure follows by replacing $D/E^2$ in Eq.~\eqref{eq:orbitalpr} with $\alpha^2L^2/(2r^2E^2)$. Equations~\eqref{eq:einstein}, \eqref{eq:actions} and \eqref{eq:orbitalrho}--\eqref{eq:orbitalpr} form a closed equilibrium problem for a specified occupation function. The energies, turning points and probabilities must be determined in the resulting metric.

Consider a differentiable family of such equilibria with a common vacuum interval outside all nearby horizons. We restrict to a locally unique $C^2$ branch for which the occupied compact support remains on regular bound tori at finite distance from the capture and separatrix boundaries. For sufficiently small parameter changes, action angle coordinates may then be chosen on a common neighbourhood of the occupied tori, with neighbouring charts related by a smooth near identity canonical transformation. An adiabatic variation preserves $\mathcal F(\mathbf J)$, or equivalently $n(J_r,L)$, while allowing the coordinate density to change. It represents a quasistatic change of the central mass on a time scale long compared with the occupied orbital periods, without capture or injection of particles. It does not identify this parameter variation with autonomous accretion of the stationary population. The fine grained Casimirs $\int\chi(f)d\Gamma$ are also preserved by Hamiltonian transport, although preserving them alone is a weaker condition than fixing the occupation of every action cell.

\subsection{Mass variation and the kinetic matter term}
\label{sec:variation}

The matter contribution to a black hole first law must be derived with the allowed variations specified \cite{Iyer1997}. For the spherical kinetic system, it can be obtained directly from the linearized mass constraint. First vary at fixed canonical $(x^i,u_i)$. Since $\sqrt q\propto B^{-1/2}$ and $\delta B=-2\delta m/r$, Eqs.~\eqref{eq:moments1}--\eqref{eq:moments2} imply
\begin{equation}
 \delta\rho=\frac1{\sqrt q}\int\varepsilon\,\delta f\,d^3u
 -\frac{\rho+p_r}{rB}\,\delta m.
 \label{eq:delta_rho}
\end{equation}
The second term includes both the variation of the proper volume and that of the local particle energy. Substitution into $\delta m'=4\pi r^2\delta\rho$ gives
\begin{equation}
 \delta m'+\frac{4\pi r(\rho+p_r)}B\,\delta m
 =4\pi r^2\frac1{\sqrt q}\int\varepsilon\,\delta f\,d^3u.
 \label{eq:mass_ode}
\end{equation}
Equation~\eqref{eq:Z} supplies its integrating factor. Integrating from a fixed sphere in the common vacuum gap to infinity, and restoring the spatial angular integration, yields
\begin{equation}
 \delta M_\infty=Z_h\,\delta M_\bullet+\int H\,\delta f\,d\Gamma.
 \label{eq:eulerian_law}
\end{equation}
The lower boundary contributes $Z_h\delta M_\bullet$ because the mass function is constant throughout the gap. The first term equals $\kappa_h\delta A_h/(8\pi)$ by Eq.~\eqref{eq:horizon}. The matter integral is therefore obtained from the constraint itself.

On the regular branch specified above, neighbouring action angle charts are related by a near identity canonical map. An admissible first variation can therefore be separated into a change of the occupation at fixed actions and a canonical displacement of the occupied tori. Locally, write
\begin{equation}
 \delta f=\left.\delta\mathcal F\right|_{\mathbf J}+\{G,f\},\qquad
 \int H\{G,f\}\,d\Gamma=\int G\{f,H\}\,d\Gamma=0.
 \label{eq:canonical_variation}
\end{equation}
Here $G$ is a smooth generator periodic in the angle coordinates. Compact support and the absence of phase space flux remove the boundary terms. The displacement integral vanishes by stationarity, rather than by setting the Eulerian $\delta f$ to zero. Equation~\eqref{eq:eulerian_law} consequently becomes
\begin{equation}
 \delta M_\infty=\frac{\kappa_h}{8\pi}\delta A_h
 +(2\pi)^3\int E(\mathbf J)\left.\delta\mathcal F\right|_{\mathbf J}d^3J
 =Z_h\delta M_\bullet+\int E\,\delta n\,dJ_r\,dL.
 \label{eq:action_law}
\end{equation}
This is the spherical kinetic form of the matter variation law. Along a family preserving the occupation function,
\begin{equation}
 \left.\frac{dM_\infty}{dM_\bullet}\right|_n=Z_h,
 \qquad 0<\left.\frac{dM_\infty}{dM_\bullet}\right|_n\leq1.
 \label{eq:constrained_slope}
\end{equation}
The total rest mass is constant on this family, but its contribution to the ADM mass need not be. At fixed $M_\bullet$, Eq.~\eqref{eq:action_law} instead assigns the energy $E$ to an infinitesimal addition of occupation. In vacuum it reduces to $dM_\infty=dM_\bullet$. These are distinct checks on the normalization and interpretation of the law.

For the locally $C^2$ equilibrium branch, let $\eta(J_r,L)$ be a smooth compactly supported occupation direction within the regular action domain, and define $D_n Z_h[\eta]=\left.dZ_h[M_\bullet,n+\epsilon\eta]/d\epsilon\right|_{\epsilon=0}$. Commuting the central mass variation with this directional occupation variation gives the weak reciprocity identity
\begin{equation}
 D_n Z_h[\eta]
 =\int \eta(J_r,L)
 \left.\frac{\partial E(J_r,L)}{\partial M_\bullet}\right|_n dJ_r\,dL .
 \label{eq:reciprocity_weak}
\end{equation}
This form is sufficient for all finite dimensional occupation modes used below and includes the response of the self gravitational field. If the directional derivative of $Z_h$ admits a kernel representation on the occupied action domain, Eq.~\eqref{eq:reciprocity_weak} is equivalently
\begin{equation}
 \left.\frac{\partial E(J_r,L)}{\partial M_\bullet}\right|_n
 =\frac{\delta Z_h}{\delta n(J_r,L)}
 \quad\hbox{almost everywhere}.
 \label{eq:reciprocity}
\end{equation}
The pointwise form is therefore a representation of the integrated identity rather than an additional assumption used by the numerical calculation.

\subsection{Dilute orbital benchmark and variation paths}
\label{sec:benchmark}

An explicit check is obtained by taking a spherical, phase mixed population of rest mass $\mu\ll M_\bullet$ concentrated on a bound Schwarzschild orbit. Averaging over orbital orientations makes the source spherical. A narrow action cell provides the corresponding regular limiting interpretation. The orbit is labelled by the dimensionless semilatus rectum $p$ and eccentricity $e$ \cite{Cutler1994}:
\begin{align}
 r(\chi)&=\frac{M_\bullet p}{1+e\cos\chi},\qquad
 E^2=\frac{(p-2)^2-4e^2}{p(p-3-e^2)},\qquad
 \frac{L^2}{M_\bullet^2}=\frac{p^2}{p-3-e^2},
 \label{eq:darwin}\\
 \frac{dt}{d\chi}&=\frac{M_\bullet p^2\sqrt{(p-2)^2-4e^2}}
 {(p-2-2e\cos\chi)(1+e\cos\chi)^2\sqrt{p-6-2e\cos\chi}}.
 \label{eq:darwin_time}
\end{align}
We use the stable bound domain $0\leq e<1$, $p>6+2e$. Set $B_0=1-2M_\bullet/r$ and $R=E^2-B_0(1+L^2/r^2)$. To first order in $\mu$, Eqs.~\eqref{eq:orbitalrho}--\eqref{eq:orbitalpr} give
\begin{equation}
 4\pi r^2\rho=\mu E P+O[(\mu/M_\bullet)^2],\qquad
 4\pi r^2p_r=\mu E P\frac{R}{E^2}+O[(\mu/M_\bullet)^2],
 \label{eq:dilute_moments}
\end{equation}
at fixed background length scales. The total mass and horizon ratio are
\begin{align}
 M_\infty&=M_\bullet+\mu E+O(\mu^2/M_\bullet),
 \label{eq:dilute_mass}\\
 Z_h&=1-\frac{\mu}{M_\bullet}K(p,e)+O[(\mu/M_\bullet)^2],
 \quad
 K=M_\bullet\left\langle\frac{E}{rB_0}
 +\frac{R}{rB_0E}\right\rangle_t.
 \label{eq:dilute_K}
\end{align}
The first and second terms in $K=K_\rho+K_{p_r}$ are respectively the density and radial pressure contributions. Comparing Eqs.~\eqref{eq:constrained_slope}, \eqref{eq:dilute_mass} and \eqref{eq:dilute_K} predicts
\begin{equation}
 K=-M_\bullet\left.\frac{\partial E}{\partial M_\bullet}\right|_{J_r,L}.
 \label{eq:orbit_check}
\end{equation}
It is essential to adjust both $(p,e)$ when differentiating an eccentric orbit at fixed dimensional actions.

\begin{figure}[!t]
\centering
\includegraphics[width=0.96\linewidth]{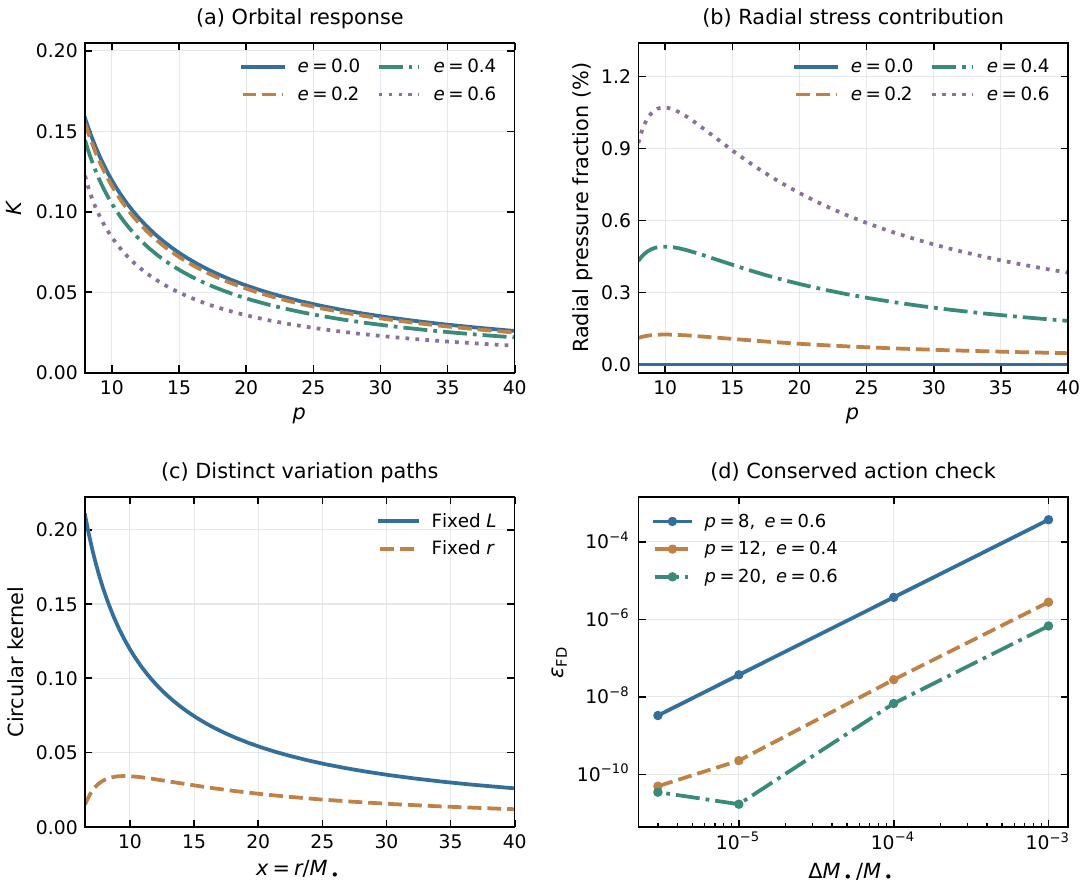}
\caption{Calculated dilute response and verification. (a) The complete kernel $K$ for $e=0,0.2,0.4,0.6$ over $8\leq p\leq40$. (b) The fraction $100K_{p_r}/K$ contributed by radial pressure. (c) Circular kernels for fixed angular momentum and fixed areal radius, defined in Eqs.~\eqref{eq:circular_K} and \eqref{eq:fixed_radius}. (d) Relative error of conserved action central differences for three representative orbits as the dimensionless mass step decreases. The common reference uses the stress integral. These calculations concern the coefficient linear in the population rest mass.}
\label{fig:response}
\end{figure}

There are two independent orbital evaluations of Eq.~\eqref{eq:orbit_check}. The first solves for neighbouring orbits with the same $(J_r,L)$ and differences their energies. The second uses Schwarzschild scaling: $E$ depends on the actions through $J_r/M_\bullet$ and $L/M_\bullet$. Since $\partial E/\partial J_r=\Omega_r$ and $\partial E/\partial L=\Omega_\phi$, it follows that
\begin{equation}
 K=J_r\Omega_r+L\Omega_\phi,\qquad
 \Omega_r=\frac{2\pi}{T_r},\quad
 \Omega_\phi=\frac{\Delta\phi}{T_r},\quad
 \Delta\phi=2\int_0^\pi\sqrt{\frac p{p-6-2e\cos\chi}}\,d\chi.
 \label{eq:scaling_check}
\end{equation}
All frequencies use the time normalized at infinity. Table~\ref{tab:orbit_checks} compares the stress integral with conserved action finite differences. The transformation in Eq.~\eqref{eq:darwin} removes the integrable turning point singularities. The table uses 256 quadrature nodes on $0\leq\chi\leq\pi$; reducing this to 64 changes $K$ by less than $10^{-13}$ in relative terms for the listed cases. The scaling expression agrees with the stress integral to better than $10^{-13}$. The finite difference comparison has a larger, explicitly resolved truncation and subtraction error, as shown in Fig.~\ref{fig:response}(d).

\begin{table}[!t]
\centering\small
\caption{Dilute orbital checks at $M_\bullet=1$. Here $K$ is calculated from the density and pressure integral in Eq.~\eqref{eq:dilute_K}, $K_{\rm FD}$ uses neighbouring masses $M_\bullet(1\pm3\times10^{-6})$ at fixed dimensional actions, and $\epsilon_{\rm FD}=|K_{\rm FD}-K|/K$.}
\label{tab:orbit_checks}
\begin{tabular}{@{}rrrrrrr@{}}
\toprule
$p$ & $e$ & $J_r/M_\bullet$ & $L/M_\bullet$ & $K$ & $K_{\rm FD}$ & $\epsilon_{\rm FD}$\\
\midrule
8 & 0.0 & 0.0000000 & 3.5777088 & 0.158113883 & 0.158113883 & $9.53\times10^{-11}$ \\
8 & 0.2 & 0.0494573 & 3.5921060 & 0.154970279 & 0.154970279 & $3.30\times10^{-10}$ \\
8 & 0.6 & 0.6282197 & 3.7139068 & 0.122369184 & 0.122369184 & $3.31\times10^{-9}$ \\
12 & 0.4 & 0.3109900 & 4.0360368 & 0.083405991 & 0.083405991 & $4.96\times10^{-11}$ \\
20 & 0.6 & 1.1306713 & 4.9029034 & 0.035724262 & 0.035724262 & $3.48\times10^{-11}$ \\
40 & 0.4 & 0.5814694 & 6.5902241 & 0.021971081 & 0.021971081 & $8.48\times10^{-10}$ \\
\bottomrule
\end{tabular}

\end{table}

Figure~\ref{fig:response}(a) gives the resulting response kernel for four eccentricities. At fixed $p$, increasing $e$ does not preserve the binding energy or either action; the curves display the dependence of the kernel on orbital shape and size, rather than an adiabatic trajectory. Panel (b) resolves the radial pressure contribution. It vanishes for circular motion and remains positive for the eccentric cases. Even when numerically small, it is required for the equality in Eq.~\eqref{eq:orbit_check}.

For a circular orbit, $x=r/M_\bullet>6$ and the same calculation is elementary:
\begin{equation}
 E_c=\frac{1-2/x}{\sqrt{1-3/x}},\qquad
 L_c^2=\frac{M_\bullet r^2}{r-3M_\bullet},\qquad
 K_L=\frac1{x\sqrt{1-3/x}}.
 \label{eq:circular_K}
\end{equation}
Holding $L_c$ fixed yields $dr/dM_\bullet=-r^2/[M_\bullet(r-6M_\bullet)]$. If the areal radius is held fixed instead, the angular momentum must change and
\begin{equation}
 K_r\equiv-M_\bullet\left.\frac{\partial E_c}{\partial M_\bullet}\right|_r
 =\frac{1-6/x}{2x(1-3/x)^{3/2}},\qquad
 \left.\frac{\partial E_c}{\partial M_\bullet}\right|_r
 =-\frac{K_L}{M_\bullet}+\Omega_\phi
 \left.\frac{\partial L_c}{\partial M_\bullet}\right|_r.
 \label{eq:fixed_radius}
\end{equation}
The extra term is angular momentum work associated with changing the orbit population. Panel (c) exhibits this difference directly. In the Newtonian limit $K_L\sim x^{-1}$ and $K_r\sim(2x)^{-1}$. As $x\to6^+$, the fixed angular momentum orbital radius derivative $dr/dM_\bullet=-r^2/[M_\bullet(r-6M_\bullet)]$ diverges, signalling the loss of the regular adiabatic branch; the kernels themselves have the limits $K_L\to\sqrt2/6$ and $K_r\to0$. All numerical benchmarks remain separated from this boundary.

Equations~\eqref{eq:action_law} and \eqref{eq:reciprocity_weak} specify the response problem for a self gravitating population. They separate changes of horizon mass from changes of orbital occupation and supply independent checks for equilibrium sequences. The dilute results in Table~\ref{tab:orbit_checks} and Fig.~\ref{fig:response} establish their orbital normalization and the distinction between the two variation prescriptions.

\section{Self gravitating populations and nonlinear horizon response}
\label{sec:selfgravity}

We now construct equilibria in which the occupied orbits and their gravitational field are determined together. Distribution functions specified by actions have an established role in self consistent stellar dynamics \cite{Posti2015}. Here the construction uses the relativistic Hamiltonian in Eq.~\eqref{eq:canonical}, includes both kinetic pressures, and retains the vacuum interval required by the horizon variation law. The occupation function specifies a controlled family of collisionless shells. Its parameters describe the particle population rather than an imposed spatial density profile.

\subsection{A compact occupation family}
\label{sec:compact_family}

Introduce a fixed reference mass $M_0$ and measure masses, radii and specific actions in units of $M_0$. This reference is independent of $M_\bullet$ throughout every sequence. We take
\begin{align}
 n(J_r,L)&=\mu h(J_r,L),\qquad
 h(J_r,L)=\frac{b[(J_r-J_c)/\Delta_J]\,b[(L-L_c)/\Delta_L]}{\Delta_J\Delta_L},
 \label{eq:compact_occupation}\\
 b(s)&=\begin{cases}\displaystyle\frac{315}{256}(1-s^2)^4,& |s|<1,\\[2pt]0,&|s|\geq1.\end{cases}
 \label{eq:bump}
\end{align}
The function $b$ is nonnegative, has three continuous derivatives, and obeys $\int b(s)\,ds=1$ and $\int s^2b(s)\,ds=1/11$. Thus $\mu=M_{\rm rest}$, the mean actions are $J_c,L_c$, and their variances are $\Delta_J^2/11,\Delta_L^2/11$. We require $J_c>\Delta_J>0$ and $L_c>\Delta_L>0$. The normalization is imposed in action space; no adjustment of $\mu$ is made after reconstructing the spatial moments.

The central mass sequences use
\begin{equation}
 (J_c,\Delta_J,L_c,\Delta_L)=(0.30,0.20,5.20,0.30)M_0,
 \quad 0.85\leq M_\bullet/M_0\leq1.15,
 \quad \mu/M_0\in\{0.10,0.30,0.50\}.
 \label{eq:reference_family}
\end{equation}
At $M_\bullet=M_0$, a second set uses $J_c/M_0\in\{0.15,0.30,0.60\}$, $\Delta_J=2J_c/3$ and $0.002\leq\mu/M_0\leq0.50$, with $L_c,\Delta_L$ unchanged. These sequences compare different mean radial actions with the same relative spread. A separate width comparison holds $J_c=0.30M_0$ fixed and uses $\Delta_J/M_0\in\{0.10,0.20,0.28\}$. This distinguishes changes of the mean radial action from changes of its spread.

\subsection{Reconstruction in local momentum variables}
\label{sec:momentum_reconstruction}

The orbital probability in Eq.~\eqref{eq:probability} has integrable turning point singularities. For the field reconstruction we instead integrate over local radial momentum,
\begin{equation}
 w=\sqrt B\,u_r,\qquad
 \varepsilon=\sqrt{1+w^2+L^2/r^2},\qquad E=\alpha\varepsilon,
 \qquad \frac{d^3u}{\sqrt q}=\frac{2\pi L}{r^2}\,dL\,dw.
 \label{eq:local_measure}
\end{equation}
The last equality includes the integration over the direction of tangential momentum. Define $g_\rho=4\pi r^2\rho$, $g_r=4\pi r^2p_r$, $g_t=4\pi r^2p_t$ and $g_0=4\pi r^2\rho_0$, where $\rho_0$ is the local rest mass density. Using Eq.~\eqref{eq:occupation} and the symmetry under $w\mapsto-w$ gives
\begin{equation}
 \begin{pmatrix}g_\rho\\g_r\\g_t\\g_0\end{pmatrix}
 =\frac1\pi\int dL\int_0^\infty dw\,
 n[J_r(E,L),L]
 \begin{pmatrix}\varepsilon\\w^2/\varepsilon\\L^2/(2r^2\varepsilon)\\1\end{pmatrix}.
 \label{eq:local_sources}
\end{equation}
The integrand is evaluated only on the occupied outer bound component; it is zero on the disconnected component that reaches the horizon. For each $L$, let $E_\pm(L)$ correspond to $J_r=J_c\pm\Delta_J$. The allowed local momentum interval is obtained from $w_\pm^2=[E_\pm^2/\alpha^2-1-L^2/r^2]_+$, together with this component restriction, where $[x]_+=\max(x,0)$.

For radii $r_b$ and $r_o$ in the inner and outer vacuum regions, the metric follows from
\begin{align}
 m(r)&=M_\bullet+\int_{r_b}^r g_\rho(s)\,ds,
 \label{eq:mass_integral}\\
 Z(r)&=\exp\!\left[-\int_r^{r_o}\frac{g_\rho(s)+g_r(s)}{sB(s)}\,ds\right],
 \qquad \alpha(r)=Z(r)\sqrt{B(r)}.
 \label{eq:lapse_integral}
\end{align}
We use $r_b=2.5M_0$ and $r_o=70M_0$. Both boundaries lie in vacuum for the computed families. The exterior beyond $r_o$ is Schwarzschild with mass $M_\infty=m(r_o)$ and $Z=1$; the inner interval has $m=M_\bullet$ and $Z=Z_h$.

Equations~\eqref{eq:actions} and \eqref{eq:local_sources}--\eqref{eq:lapse_integral} are iterated to consistency. Each action inversion uses $\partial J_r/\partial E=T_r/(2\pi)$ and a bracket restricted to the outer potential well. The transformation $r=(r_a+r_p)/2-(r_a-r_p)\cos\chi/2$, with $0\leq\chi\leq\pi$, regularizes the orbital integrals. Polynomial interpolation between action nodes then supplies $J_r(E,L)$ for the local momentum integrals. The inner barrier used in the orbital bracket lies in the vacuum gap of the retained configurations. There the constant lapse ratio leaves the Schwarzschild barrier radius unchanged. The stable outer orbit and both turning points are evaluated in the complete metric. The standard resolution is $(N_r,N_L,N_w,N_J,N_\chi)=(1601,28,36,20,96)$: a uniform radial grid, Gaussian angular and radial momentum quadratures, action interpolation nodes, and Gaussian orbital quadrature. Iteration stops when the largest difference between reconstructed and current values of $m/M_0$ and $Z$ is below $5\times10^{-11}$. This tolerance controls the iteration; the discretization and derivative errors are assessed separately below.

\subsection{Mass accounting and occupation derivatives}
\label{sec:mass_accounting}

Three integral quantities provide distinct checks. The reconstructed rest mass is
\begin{equation}
 \mu_{\rm rec}=\int_{r_b}^{r_o}\frac{g_0(r)}{\sqrt{B(r)}}\,dr.
 \label{eq:rest_check}
\end{equation}
It must equal the prescribed action occupation $\mu$. The Komar mass associated with the Killing field normalized at infinity \cite{Komar1959} is
\begin{align}
 M_K(r)&=r^2\sqrt B\,\alpha'=Z(m+r g_r),
 \label{eq:komar_local}\\
 M_\infty&=Z_hM_\bullet+\int_{r_b}^{r_o} Z(g_\rho+g_r+2g_t)\,dr.
 \label{eq:komar_global}
\end{align}
The local expression follows from Eq.~\eqref{eq:einstein}; its derivative is $M_K'=Z(g_\rho+g_r+2g_t)$. The inner boundary term is $Z_hM_\bullet=\kappa_h A_h/(4\pi)$, and the exterior value is $M_\infty$.

Finally, define the mean specific Killing energy of the population by
\begin{equation}
 \overline E(M_\bullet,\mu)=\int h(J_r,L)E(J_r,L;M_\bullet,\mu)\,dJ_r\,dL,
 \qquad \mu\overline E=\int_{r_b}^{r_o} Zg_\rho\,dr.
 \label{eq:mean_energy}
\end{equation}
The second equality follows by integrating Eq.~\eqref{eq:orbitalrho} and using $\int P\,dr=1$. Subtracting it from Eq.~\eqref{eq:mass_integral} yields
\begin{equation}
 \mathcal W\equiv M_\infty-M_\bullet-\mu\overline E
 =\int_{r_b}^{r_o}(1-Z)g_\rho\,dr\geq0.
 \label{eq:energy_defect}
\end{equation}
This identity specifies the difference between the matter contribution to the ADM mass and the sum of its particle Killing energies. The factor $Z$ accounts for the gravitational weighting of those energies. It is distinct from the binding energy defined relative to $M_\bullet+\mu$. For a regular branch connected to vacuum, $1-Z=O(\mu/M_0)$ and $g_\rho=O(\mu/M_0)$ at fixed length scales, so $\mathcal W=O(\mu^2/M_0)$.

For variations at fixed shape $h$, Eq.~\eqref{eq:action_law} gives
\begin{equation}
 \left.\frac{\partial M_\infty}{\partial M_\bullet}\right|_{\mu,h}=Z_h,
 \qquad
 \left.\frac{\partial M_\infty}{\partial\mu}\right|_{M_\bullet,h}=\overline E,
 \qquad
 \left.\frac{\partial\overline E}{\partial M_\bullet}\right|_{\mu,h}
 =\left.\frac{\partial Z_h}{\partial\mu}\right|_{M_\bullet,h}.
 \label{eq:amplitude_identities}
\end{equation}
The last relation tests the projection of Eq.~\eqref{eq:reciprocity_weak} along a change of occupation amplitude. All derivatives include the induced change of the metric and the orbits. Equivalently, integration along a branch starting in vacuum gives
\begin{equation}
 M_\infty(M_\bullet,\mu)=M_\bullet+\int_0^\mu\overline E(M_\bullet,\eta)\,d\eta,
 \quad
 \mathcal W=\int_0^\mu[\overline E(M_\bullet,\eta)-\overline E(M_\bullet,\mu)]\,d\eta.
 \label{eq:amplitude_integral}
\end{equation}
Thus adding the final particle energies to the horizon mass omits a term already present at second order in the occupation amplitude.

\subsection{Equilibrium sequences and nonlinear response}
\label{sec:nonlinear_results}

\begin{figure}[!t]
\centering
\includegraphics[width=0.97\linewidth]{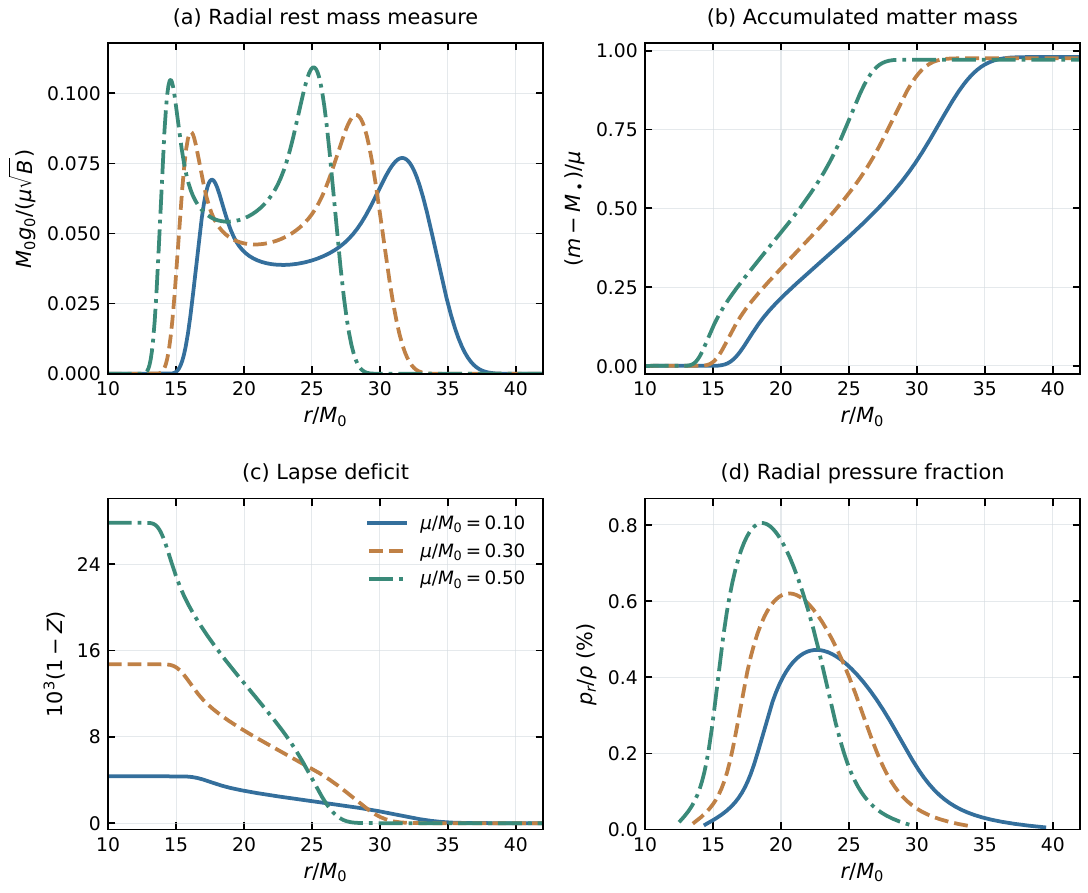}
\caption{Computed equilibrium profiles for $M_\bullet=M_0$ and the occupation shape in Eq.~\eqref{eq:reference_family}. The three curves use $\mu/M_0=0.10,0.30,0.50$. (a) The radial rest mass density divided by the prescribed total rest mass, expressed in the dimensionless coordinate $r/M_0$. (b) The accumulated matter contribution to the mass function divided by $\mu$. Its exterior value is $(M_\infty-M_\bullet)/\mu$. (c) The lapse deficit $10^3(1-Z)$, including its constant value in the inner vacuum interval. (d) Radial pressure relative to energy density, shown where $g_\rho$ exceeds $10^{-4}$ of its maximum in the corresponding configuration. All profiles are reconstructed from the action occupation.}
\label{fig:profiles}
\end{figure}

Figure~\ref{fig:profiles} shows the reconstructed profiles at $M_\bullet=M_0$. The radial rest mass measure is $g_0\,dr/\sqrt B$, and we define its median by
\begin{equation}
 \int_{r_b}^{r_{50}}\frac{g_0}{\sqrt B}\,dr=\frac12\mu_{\rm rec}.
 \label{eq:median_radius}
\end{equation}
Increasing the occupation amplitude at fixed actions contracts the population and increases its effect on the horizon normalization. For the reference shape, $r_{50}/M_0$ decreases from 26.9374 at $\mu=0.10M_0$ to 21.1968 at $\mu=0.50M_0$, while $Z_h$ decreases from 0.99565132 to 0.97217207. The radial pressure follows from the same eccentric orbits. Its share of the integral $-\ln Z_h$, defined by the ratio of $\int g_r/(rB)\,dr$ to $\int(g_\rho+g_r)/(rB)\,dr$, increases from 0.241 to 0.419 per cent over these two configurations.

\begin{figure}[!t]
\centering
\includegraphics[width=0.97\linewidth]{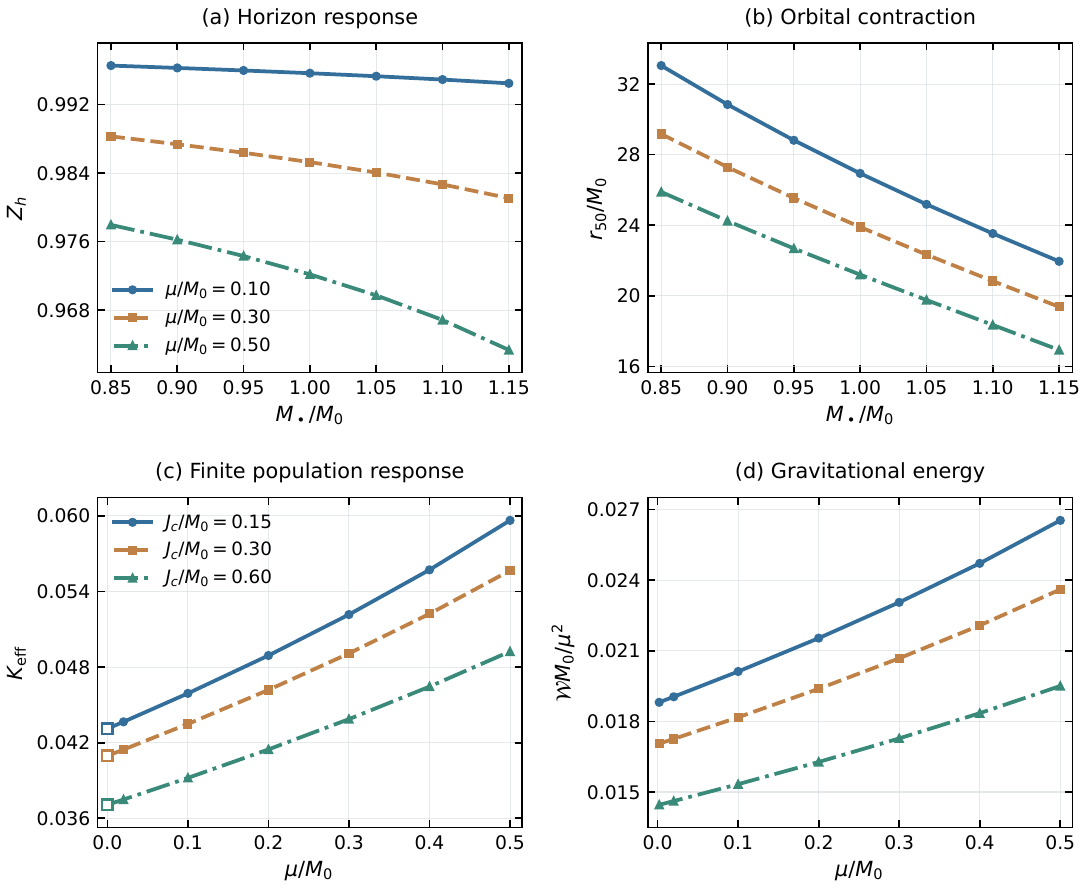}
\caption{Response of the computed equilibrium families. (a,b) Horizon lapse ratio and median rest mass radius along central mass sequences with fixed dimensional action occupations. (c) Effective response coefficient as a function of occupation amplitude at $M_\bullet=M_0$, for three mean radial actions with $\Delta_J=2J_c/3$. Open squares at $\mu=0$ show independent Schwarzschild averages $K_0$. (d) The positive energy difference $\mathcal W M_0/\mu^2$, evaluated from the radial integral in Eq.~\eqref{eq:energy_defect}, for the same three families.}
\label{fig:nonlinear}
\end{figure}

Selected results along the central mass sequences are given in Table~\ref{tab:equilibria}, and their response is shown in Fig.~\ref{fig:nonlinear}. At fixed occupation, increasing $M_\bullet$ makes the matter distribution more compact and decreases $Z_h$ throughout the sampled interval. For example, at $\mu=0.30M_0$, increasing $M_\bullet/M_0$ from $0.85$ to $1.15$ changes $r_{50}/M_0$ from 29.1694 to 19.3744. The constrained derivative of the matter contribution satisfies $\partial(M_\infty-M_\bullet)/\partial M_\bullet=Z_h-1<0$: the matter contribution decreases as the central mass increases, even though the population rest mass is preserved.

\begin{table}[!t]
\centering
\small

\caption{Selected equilibria with the common action occupation shape in Eq.~\eqref{eq:reference_family}. The energy difference $\mathcal W$ uses the radial integral in Eq.~\eqref{eq:energy_defect}. All masses and radii are expressed using the fixed reference $M_0$.}
\label{tab:equilibria}
\begin{tabular}{rrrrrr}
\toprule
$M_\bullet/M_0$ & $\mu/M_0$ & $M_\infty/M_0$ & $Z_h$ & $r_{50}/M_0$ & $10^3\mathcal W/M_0$ \\
\midrule
0.85 & 0.10 & 0.94857733 & 0.99654314 & 33.0474 & 0.14644 \\
1.00 & 0.10 & 1.09799447 & 0.99565132 & 26.9374 & 0.18175 \\
1.15 & 0.10 & 1.24725856 & 0.99446984 & 21.9521 & 0.22675 \\
0.85 & 0.30 & 1.14479274 & 0.98827910 & 29.1694 & 1.50260 \\
1.00 & 0.30 & 1.29281941 & 0.98527529 & 23.8944 & 1.86133 \\
1.15 & 0.30 & 1.44031688 & 0.98106721 & 19.3744 & 2.34639 \\
0.85 & 0.50 & 1.33953993 & 0.97794775 & 25.8845 & 4.74402 \\
1.00 & 0.50 & 1.48582220 & 0.97217207 & 21.1968 & 5.90188 \\
1.15 & 0.50 & 1.63104593 & 0.96334791 & 16.9287 & 7.61634 \\
\bottomrule
\end{tabular}
\end{table}

To compare with the dilute response, define
\begin{equation}
 K_{\rm eff}=\frac{M_\bullet}{\mu}(1-Z_h),\qquad
 K_0=\int h(J_r,L)K(J_r,L;M_\bullet)\,dJ_r\,dL
 =\lim_{\mu\to0}K_{\rm eff},
 \label{eq:effective_response}
\end{equation}
where the kernel in $K_0$ is evaluated in the Schwarzschild background using Eq.~\eqref{eq:dilute_K}. For $M_\bullet=M_0$ and the reference shape, independent orbital quadrature gives $K_0=0.04095724$. At $\mu=0.50M_0$, $K_{\rm eff}=0.05565586$, an increase of 35.89 per cent. This is the effect of the population's gravitational field on both the source and its orbital support. The three sequences in Fig.~\ref{fig:nonlinear}(c) approach their separately calculated dilute limits.

The fixed mean width comparison gives a smaller effect over the chosen interval. At $(M_\bullet,\mu,J_c)=(1,0.50,0.30)M_0$, increasing $\Delta_J/M_0$ from $0.10$ to $0.28$ changes $K_{\rm eff}$ from 0.05565057 to 0.05566426, or 0.0246 per cent, while $r_{50}/M_0$ changes from 21.2371 to 21.1300. This result concerns the stated compact family and separates its width dependence from the larger differences among the mean action sequences.

The positive energy difference in Eq.~\eqref{eq:energy_defect} is resolved independently of these comparisons. For the reference shape at $(M_\bullet,\mu)=(1,0.50)M_0$, $\mathcal W/M_0=5.90188\times10^{-3}$. Figure~\ref{fig:nonlinear}(d) displays its quadratic scaling normalization across the occupation sequences. The radial integral is used for this plot because it avoids subtracting nearly equal masses in the smallest amplitude configurations. The direct mass and orbital energy expression is retained as a separate consistency check.

\subsection{Conservation, resolution and derivative checks}
\label{sec:nonlinear_checks}

The calculation comprises 73 completed equilibrium runs, including the sequence points, neighboring configurations for derivative tests and four resolutions. At the standard resolution, the largest relative discrepancies are $4.21\times10^{-8}$ in Eq.~\eqref{eq:rest_check}, $2.90\times10^{-10}$ in Eq.~\eqref{eq:komar_global}, and $4.21\times10^{-8}$ in Eq.~\eqref{eq:mean_energy}. The largest absolute discrepancy between the two evaluations of $\mathcal W/M_0$ is $1.23\times10^{-8}$. These checks compare separately reconstructed quantities without correcting the occupation normalization.

Table~\ref{tab:resolution} gives the resolution study at $(M_\bullet,\mu)=(1,0.30)M_0$. We define $\epsilon_\mu=|\mu_{\rm rec}/\mu-1|$, $\epsilon_K=|M_K(r_o)/M_\infty-1|$ using the integrated expression in Eq.~\eqref{eq:komar_global}, and $\epsilon_Z=|Z_h/Z_h^{\rm fine}-1|$. The derivative discrepancy is $\epsilon_M=|D_{M_\bullet}M_\infty/Z_h-1|$, where
\begin{equation}
 D_x F=\frac{F(x+hM_0)-F(x-hM_0)}{2hM_0},
 \qquad x\in\{M_\bullet,\mu\}.
 \label{eq:central_difference}
\end{equation}
The finest calculation gives $\epsilon_M=6.67\times10^{-9}$ for $h=10^{-3}$. The changes of $M_\infty$ and $Z_h$ between the standard and finest reference configurations are respectively $4.84\times10^{-10}$ and $6.63\times10^{-11}$ in relative terms. These integral comparisons have a different accuracy from derivatives of the interpolated radial fields. For example, the normalized residual $\max|m'-g_\rho|/\max g_\rho$ decreases from $6.52\times10^{-4}$ at the coarsest resolution to $4.42\times10^{-7}$ at the finest.

Local stress conservation provides an additional check involving the independently reconstructed tangential pressure:
\begin{equation}
 g_r'+(g_\rho+g_r)\frac{\alpha'}{\alpha}-\frac{2g_t}{r}=0.
 \label{eq:local_stress_balance}
\end{equation}
This is the radial conservation equation after multiplication by $4\pi r^2$. Let $F$ denote its left hand side. We use $\epsilon_F=\max|F|/\max\{\max|g_r'|,\max|(g_\rho+g_r)\alpha'/\alpha|,\max|2g_t/r|\}$. Both derivatives are obtained from the radial interpolants of $g_r$ and $\ln\alpha$, without substituting the field equation for $\alpha'/\alpha$. Table~\ref{tab:resolution} shows the resulting convergence. Across all standard resolution configurations in this section, $\epsilon_F\leq 1.88\times10^{-6}$.

\begin{table}[!t]
\centering
\small

\setlength{\tabcolsep}{4pt}
\caption{Resolution checks at $(M_\bullet,\mu)=(1,0.30)M_0$. The tuple lists $(N_r,N_L,N_w,N_J,N_\chi)$. The standard resolution is the third row. The finest row supplies the reference for $\epsilon_Z$; the dash denotes that reference. The mass slope uses $h=10^{-3}$ at each resolution. The final column tests Eq.~\eqref{eq:local_stress_balance}.}
\label{tab:resolution}
\begin{tabular}{@{}lrrrrr@{}}
\toprule
Resolution & $\epsilon_\mu$ & $\epsilon_K$ & $\epsilon_Z$ & $\epsilon_M$ & $\epsilon_F$ \\
\midrule
$(401,12,16,12,48)$ & $1.29\times10^{-6}$ & $7.77\times10^{-9}$ & $3.29\times10^{-8}$ & $4.07\times10^{-4}$ & $6.59\times10^{-5}$ \\
$(801,20,24,16,64)$ & $6.12\times10^{-10}$ & $2.01\times10^{-11}$ & $5.19\times10^{-11}$ & $6.07\times10^{-7}$ & $3.05\times10^{-6}$ \\
$(1601,28,36,20,96)$ & $2.16\times10^{-9}$ & $1.67\times10^{-11}$ & $6.63\times10^{-11}$ & $6.95\times10^{-7}$ & $1.22\times10^{-7}$ \\
$(2401,36,48,24,128)$ & $1.10\times10^{-11}$ & $6.51\times10^{-14}$ & \textemdash & $6.67\times10^{-9}$ & $2.32\times10^{-8}$ \\
\bottomrule
\end{tabular}
\end{table}

The three relations in Eq.~\eqref{eq:amplitude_identities} are tested separately in Table~\ref{tab:derivative_checks}. We use $\epsilon_\mu^{\rm slope}=|D_\mu M_\infty/\overline E-1|$ and $\epsilon_\times=|D_{M_\bullet}\overline E-D_\mu Z_h|/|D_\mu Z_h|$. At $h=10^{-3}$, the mixed derivatives are $M_0D_{M_\bullet}\overline E=-0.058113530$ and $M_0D_\mu Z_h=-0.058113495$. Their relative discrepancy is $6.10\times10^{-7}$. Decreasing the step eventually amplifies the spatial quadrature error in mass differences, as the first column of discrepancies shows; the finest resolution result in Table~\ref{tab:resolution} supplies the corresponding independent check. The mixed derivative test addresses changes proportional to $h(J_r,L)$, rather than arbitrary localized variations in action space.

\begin{table}[!t]
\centering
\small

\caption{Central difference tests of Eq.~\eqref{eq:amplitude_identities} at $(M_\bullet,\mu)=(1,0.30)M_0$, using the standard resolution. Each neighboring configuration is solved at the same dimensional action shape. The discrepancies compare mass derivatives with $Z_h$ and $\overline E$, and compare the two mixed derivatives.}
\label{tab:derivative_checks}
\begin{tabular}{rrrr}
\toprule
$h$ & $\epsilon_M$ & $\epsilon_\mu^{\rm slope}$ & $\epsilon_\times$ \\
\midrule
$1.00\times10^{-2}$ & $8.04\times10^{-7}$ & $8.13\times10^{-7}$ & $4.73\times10^{-5}$ \\
$3.00\times10^{-3}$ & $3.11\times10^{-7}$ & $1.66\times10^{-7}$ & $3.61\times10^{-6}$ \\
$1.00\times10^{-3}$ & $6.95\times10^{-7}$ & $3.22\times10^{-7}$ & $6.10\times10^{-7}$ \\
$3.00\times10^{-4}$ & $1.01\times10^{-6}$ & $3.41\times10^{-7}$ & $4.64\times10^{-7}$ \\
\bottomrule
\end{tabular}
\end{table}

Additional action checks use interpolation evaluation points absent from the fitted node set and twice as many orbital quadrature nodes. The maximum relative action discrepancies are $9.15\times10^{-10}$ for the reference configuration, $5.70\times10^{-9}$ for $(M_\bullet,\mu)=(1.15,0.50)M_0$, and $8.07\times10^{-10}$ for $(M_\bullet,\mu,J_c,\Delta_J)=(1,0.50,0.60,0.40)M_0$. Evaluations at the four corners of the action support in the second configuration give a minimum pericentre of $9.284M_0$, compared with $r_h=2.30M_0$, and a minimum difference of $0.05136$ between the inner potential barrier energy and those corner energies. These checks verify the separation of the sampled orbital support from capture. The largest apocentre among the orbit nodes in all runs is $58.02M_0$, inside the exterior matching radius. The static families therefore provide controlled tests of the constrained response identities over the reported parameter range. Dynamical stability is a separate question from the equilibrium and variation conditions tested here.

\section{Occupation redistribution under mass constraints}
\label{sec:redistribution}

The dependence of horizon normalization on surrounding matter is established in the thermodynamics of spherical black holes \cite{Visser1992} and in explicit environmental geometries \cite{Cardoso2022}. We consider the more restrictive question of whether this normalization can change when the horizon mass, particle rest mass and ADM mass are all preserved. The action occupation makes these constraints precise. It also supplies orbital expressions for the derivatives required to construct such families.

\subsection{Redistributions that preserve the particle rest mass}
\label{sec:redistribution_family}

Let $h_0$ denote the reference shape in Eqs.~\eqref{eq:compact_occupation}--\eqref{eq:reference_family}. Set $x=(J_r-J_c)/\Delta_J$ and $y=(L-L_c)/\Delta_L$. We use the family
\begin{equation}
 n_{\boldsymbol a}(J_r,L)=\mu h_0(J_r,L)
 \left[1+a_J Q_J(x,y)+a_L Q_L(x,y)+a_W Q_W(x,y)\right],
 \label{eq:redistribution_occupation}
\end{equation}
with
\begin{equation}
 Q_J=x,\qquad Q_L=y,\qquad Q_W=\frac{11x^2-1}{10},
 \qquad |a_J|+|a_L|+|a_W|<1.
 \label{eq:redistribution_modes}
\end{equation}
Each $Q_i$ has zero mean under $h_0$, and $|Q_i|\leq1$ on the support. The stated coefficient bound therefore guarantees positivity while preserving the same compact action support and the normalization $\int n_{\boldsymbol a}\,dJ_r\,dL=\mu$. The three modes are mutually orthogonal with squared norms $1/11,1/11,1/65$ under $h_0$. They have the following exact effects on the action moments:
\begin{align}
 \langle J_r\rangle_{\boldsymbol a}&=J_c+\frac{\Delta_J a_J}{11},\qquad
 \langle L\rangle_{\boldsymbol a}=L_c+\frac{\Delta_L a_L}{11},
 \label{eq:redistribution_means}\\
 {\rm Var}_{\boldsymbol a}(J_r)&=\Delta_J^2
 \left(\frac1{11}+\frac{2a_W}{143}-\frac{a_J^2}{121}\right),\qquad
 {\rm Var}_{\boldsymbol a}(L)=\Delta_L^2
 \left(\frac1{11}-\frac{a_L^2}{121}\right).
 \label{eq:redistribution_variances}
\end{align}
Thus the $W$ direction changes the radial action variance at fixed mean actions. The $J$ and $L$ directions change the mean radial action and the mean magnitude of angular momentum, respectively. These parameters compare distinct stationary populations. Preserving the total rest mass is weaker than preserving the occupation of every action cell or all Vlasov Casimirs; a dynamical mechanism that realizes the redistribution is not assumed in this equilibrium comparison.

The local moments are reconstructed from Eq.~\eqref{eq:local_sources} with $n=n_{\boldsymbol a}$. The gravitational field and the action to energy relation are recomputed for every coefficient choice. Neither a coordinate density profile nor a pressure prescription is held fixed.

\subsection{Orbital gradients and mixed responses}
\label{sec:shape_gradients}

Define the orbital energy gradients
\begin{equation}
 \mathcal A_i=\mu\int h_0 Q_i E\,dJ_r\,dL,
 \qquad i\in\{J,L,W\}.
 \label{eq:shape_energy_gradient}
\end{equation}
The weight in this equation is the undeformed $h_0 Q_i$, since $\partial n_{\boldsymbol a}/\partial a_i=\mu h_0Q_i$. The energy $E$ is evaluated in the deformed equilibrium. The first law in Eq.~\eqref{eq:action_law} then reads
\begin{equation}
 dM_\infty=Z_h\,dM_\bullet+\overline E_{\boldsymbol a}\,d\mu
 +\sum_i\mathcal A_i\,da_i,
 \qquad
 \overline E_{\boldsymbol a}=\frac1\mu\int n_{\boldsymbol a}E\,dJ_r\,dL.
 \label{eq:finite_shape_law}
\end{equation}
In particular, $\mathcal A_i=\partial M_\infty/\partial a_i$ at fixed $M_\bullet,\mu$ and the other coefficients. Commuting derivatives on a regular twice differentiable branch gives
\begin{equation}
 \mathcal R_i\equiv\frac{\partial Z_h}{\partial a_i}
 =\frac{\partial\mathcal A_i}{\partial M_\bullet},\qquad
 \mathcal H_{ij}\equiv\frac{\partial\mathcal A_i}{\partial a_j}
 =\mathcal H_{ji}.
 \label{eq:shape_reciprocity}
\end{equation}
All quantities in Eq.~\eqref{eq:shape_reciprocity} include the field response. The matrix $\mathcal H$ is the Hessian of the equilibrium mass with respect to occupation coefficients. Its symmetry is an integrability check, rather than a criterion for collective dynamical stability.

Table~\ref{tab:shape_gradients} compares the independently integrated $\mathcal A_i$ with central differences of the ADM mass, and compares $\mathcal R_i$ with mass derivatives of the orbital gradients. The reference is $(M_\bullet,\mu)=(1,0.30)M_0$ at $\boldsymbol a=0$. The radial and angular mean directions have positive energy gradients and positive horizon responses. The variance direction has negative gradients of smaller magnitude for this population. These signs refer to the modes in Eq.~\eqref{eq:redistribution_modes}; they are not universal signs for arbitrary occupation changes.

\begin{table}[!t]
\centering\small
\caption{Occupation gradients and reciprocal responses at $(M_\bullet,\mu)=(1,0.30)M_0$ and $\boldsymbol a=0$. Here $\epsilon_{A,i}=|D_{a_i}M_\infty/\mathcal A_i-1|$ and $\epsilon_{R,i}=|D_{M_\bullet}\mathcal A_i/\mathcal R_i-1|$, where $D$ denotes the central difference. The coefficient step is $0.02$ and the mass step is $0.001M_0$. Orbital integration supplies $\mathcal A_i$; central differences of $Z_h$ supply $\mathcal R_i$.}
\label{tab:shape_gradients}
\begin{tabular}{@{}lrrrr@{}}
\toprule
Mode & $\mathcal A_i/M_0$ & $\mathcal R_i$ & $\epsilon_{A,i}$ & $\epsilon_{R,i}$\\
\midrule
$J$ & $5.48\times10^{-5}$ & $1.05\times10^{-4}$ & $1.51\times10^{-8}$ & $4.62\times10^{-7}$ \\
$L$ & $8.23\times10^{-5}$ & $2.03\times10^{-4}$ & $2.31\times10^{-8}$ & $1.79\times10^{-6}$ \\
$W$ & $-1.99\times10^{-7}$ & $-4.38\times10^{-7}$ & $1.20\times10^{-5}$ & $4.47\times10^{-6}$ \\
\bottomrule
\end{tabular}

\end{table}

At the same reference, the computed Hessian is
\begin{equation}
 \frac{\mathcal H}{M_0}\simeq10^{-7}
 \begin{pmatrix}
-6.26112 & -4.09139 & 0.13729 \\
-4.09139 & -6.73922 & 0.04074 \\
0.13729 & 0.04074 & -0.04676
\end{pmatrix},
 \label{eq:numerical_shape_hessian}
\end{equation}
in the order $(J,L,W)$. The relative antisymmetric residual $\|\mathcal H-\mathcal H^T\|_F/\|\mathcal H\|_F$ is $1.16\times10^{-8}$, evaluated before rounding the entries in Eq.~\eqref{eq:numerical_shape_hessian}. Here $\|\cdot\|_F$ denotes the Frobenius norm. The matrix is obtained from separate neighboring equilibria in each coefficient direction, without imposing symmetry on the calculated entries.

For twelve additional configurations with a single nonzero coefficient $a_i=\pm0.40,\pm0.80$, the quadratic prediction $\Delta M_\infty=\mathcal A_i a_i+\mathcal H_{ii}a_i^2/2$ differs from the calculated mass change by at most $7.57\times10^{-10}M_0$.

\subsection{A curve with fixed horizon mass, rest mass and ADM mass}
\label{sec:constant_mass_curve}

Fix $M_\bullet$ and $\mu$, set $a_W=0$, and write $a_J=a$ and $a_L=\ell(a)$. We impose $M_\infty[a,\ell(a)]=M_\infty(0,0)$. Wherever $\mathcal A_L\ne0$, the implicit function theorem gives a local curve with
\begin{equation}
 \ell'(a)=-\frac{\mathcal A_J}{\mathcal A_L},\qquad
 \ell''(a)=-\frac{\mathcal H_{JJ}+2\ell'\mathcal H_{JL}
 +(\ell')^2\mathcal H_{LL}}{\mathcal A_L}.
 \label{eq:constant_mass_tangent}
\end{equation}
The coefficients on the right are evaluated on the curve. Its horizon response is
\begin{equation}
 \mathcal S\equiv\left.\frac{dZ_h}{da}\right|_{M_\bullet,\mu,M_\infty}
 =\mathcal R_J-\frac{\mathcal A_J}{\mathcal A_L}\mathcal R_L.
 \label{eq:constant_mass_horizon_response}
\end{equation}
Consequently, a nonzero $\mathcal A_L\mathcal R_J-\mathcal A_J\mathcal R_L$ permits a change in horizon normalization while all three masses are preserved. The corresponding surface gravity ratio is exact:
\begin{equation}
 \frac{\kappa_h(a)}{\kappa_h(0)}=\frac{Z_h(a)}{Z_h(0)},\qquad
 A_h(a)=16\pi M_\bullet^2.
 \label{eq:constant_area_surface_gravity}
\end{equation}
This comparison uses the same Killing time normalization at infinity for every member of the family.

At the reference configuration, the gradients give $\ell'(0)=-0.66636001$, $\ell''(0)=0.00461864$ and $\mathcal S(0)=-3.0251517\times10^{-5}$. We solve the constant ADM mass condition directly at nine values of $a$ between $-0.50$ and $0.50$. The orbital gradient $\mathcal A_L$ supplies the correction to $\ell$ during this scalar constraint solve; the stopping condition is the actual residual in $M_\infty$. The largest absolute mass residual on this sequence is $1.86\times10^{-13}M_0$.

\begin{figure}[!t]
\centering
\includegraphics[width=0.97\linewidth]{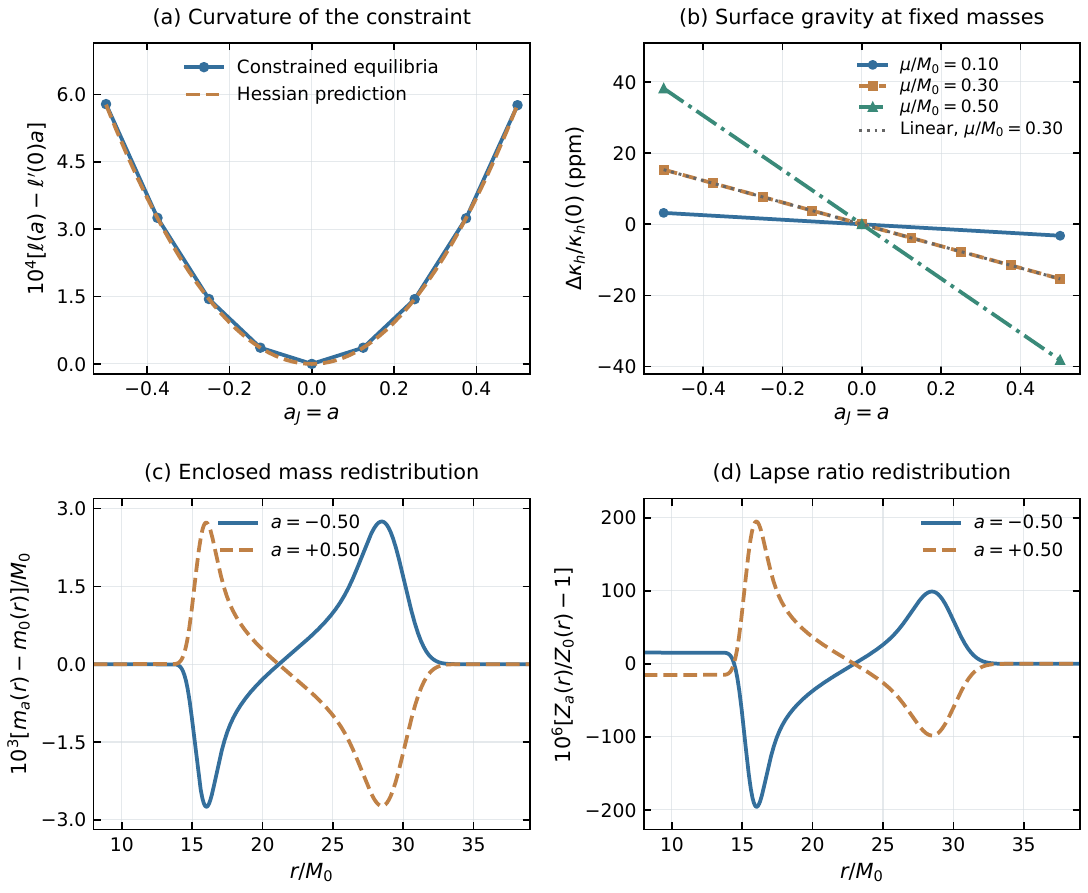}
\caption{Computed occupation redistributions at fixed $(M_\bullet,\mu,M_\infty)$ with $M_\bullet=M_0$. (a) The departure of $\ell(a)$ from its tangent at the origin for $\mu=0.30M_0$. The dashed curve uses $\ell''(0)a^2/2$ from the independently calculated Hessian. (b) Relative change in surface gravity for the same nine configurations. Additional endpoint results are shown for $\mu/M_0=0.10,0.50$; the dashed reference is the linear prediction $a\mathcal S(0)/Z_h(0)$ at $\mu=0.30M_0$. (c,d) Differences of the enclosed mass and lapse ratio relative to the undeformed population, for the two endpoint configurations at $\mu=0.30M_0$. Their outer vacuum fields coincide within the mass constraint residual, while the inner vacuum lapse ratios differ.}
\label{fig:constant_mass}
\end{figure}

Figure~\ref{fig:constant_mass} compares the resulting curve with its local derivatives and displays the corresponding radial field differences. Table~\ref{tab:constant_mass} gives the endpoints and reference configurations for $\mu/M_0=0.10,0.30,0.50$. Each group has its own fixed ADM mass. At $\mu=0.30M_0$, the surface gravity differs from the reference by $+15.37147$ parts per million at $a=-0.50$ and $-15.33219$ parts per million at $a=0.50$. Positive $a$ increases the mean radial action and decreases the mean angular momentum along this curve. The radial pressure contribution to $-\ln Z_h$ changes from 0.31125 to 0.32995 per cent between these endpoints.

\begin{table}[!t]
\centering\small
\caption{Selected configurations with $M_\bullet=M_0$ and $a_W=0$. Each rest mass group has its own fixed ADM mass. The surface gravity shift is measured relative to its undeformed member. Displayed digits retain the equality of the masses at the accuracy relevant to the response; the actual mass constraint residuals are evaluated before rounding.}
\label{tab:constant_mass}
\begin{tabular}{@{}rrrrrr@{}}
\toprule
$\mu/M_0$ & $a$ & $\ell(a)$ & $M_\infty/M_0$ & $Z_h$ & $10^6\Delta\kappa_h/\kappa_h(0)$\\
\midrule
0.10 & -0.50 & +0.3041745 & 1.0979944720 & 0.9956545191 & +3.2140 \\
0.10 & +0.00 & +0.0000000 & 1.0979944720 & 0.9956513191 & +0.0000 \\
0.10 & +0.50 & -0.3037872 & 1.0979944720 & 0.9956481249 & -3.2081 \\
\addlinespace
0.30 & -0.50 & +0.3337585 & 1.2928194049 & 0.9852904375 & +15.3715 \\
0.30 & +0.00 & +0.0000000 & 1.2928194049 & 0.9852752924 & +0.0000 \\
0.30 & +0.50 & -0.3326038 & 1.2928194049 & 0.9852601860 & -15.3322 \\
\addlinespace
0.50 & -0.50 & +0.3579212 & 1.4858221985 & 0.9722091957 & +38.1907 \\
0.50 & +0.00 & +0.0000000 & 1.4858221985 & 0.9721720678 & +0.0000 \\
0.50 & +0.50 & -0.3560312 & 1.4858221985 & 0.9721349830 & -38.1463 \\
\bottomrule
\end{tabular}

\end{table}

All three families obey the occupation positivity bound. The smallest lower bound $1-|a_J|-|a_L|$ on the constant mass sequences is 0.142079. Their occupied orbit nodes remain separated from the horizon and the outer matching radius. Since the common exterior beyond the matter is Schwarzschild with the same $M_\infty$ within each group, the differences in Fig.~\ref{fig:constant_mass}(c,d) arise inside the matter region and in its inner vacuum interval. The response therefore distinguishes the orbital content of the population under stronger constraints than a comparison at equal particle rest mass alone.

\subsection{Resolution checks and the dilute limit}
\label{sec:redistribution_checks}

The standard calculations in this section use $(N_r,N_L,N_w,N_J,N_\chi)=(2401,36,48,24,128)$. The action inversion tolerance is $5\times10^{-13}M_0$, and the equilibrium reconstruction tolerance is $2\times10^{-12}$ in $m/M_0$ and $Z$. The derivatives use dimensionless coefficient steps $\delta_a$ and central mass steps $\delta_M M_0$. Table~\ref{tab:shape_steps} reports the dependence of the three checks on the coefficient step. The reciprocal comparison uses $\delta_M=10^{-3}$; the additional $\delta_M=3\times10^{-3}$ calculation checks the mass step dependence.

\begin{table}[!t]
\centering\small
\caption{Coefficient step dependence at the standard resolution and the undeformed reference. The errors are defined in Table~\ref{tab:shape_gradients}; the mass step is $0.001M_0$. The largest relative discrepancies occur in the variance direction, whose energy gradient is smaller than those of the mean directions.}
\label{tab:shape_steps}
\begin{tabular}{@{}rrrr@{}}
\toprule
$\delta_a$ & $\max_i\epsilon_{A,i}$ & $\max_i\epsilon_{R,i}$ & $\|\mathcal H-\mathcal H^T\|_F/\|\mathcal H\|_F$\\
\midrule
0.050 & $9.28\times10^{-6}$ & $2.36\times10^{-6}$ & $1.97\times10^{-8}$ \\
0.020 & $1.20\times10^{-5}$ & $4.47\times10^{-6}$ & $1.16\times10^{-8}$ \\
0.008 & $2.39\times10^{-5}$ & $1.39\times10^{-5}$ & $3.54\times10^{-8}$ \\
\bottomrule
\end{tabular}

\end{table}

The smallest gradient occurs in the variance direction. Its relative difference errors increase at the smallest coefficient step, where the equilibrium residual is amplified by differencing. Changing the mass step from $0.003$ to $0.001$ changes $\partial\mathcal A_i/\partial M_\bullet$ by at most $1.43\times10^{-5}$ in relative terms.

A finer calculation uses $(3201,44,56,28,160)$ and an equilibrium tolerance of $10^{-12}$. At coefficient step $0.02$, its largest relative mass gradient discrepancy is $2.86\times10^{-6}$, its largest reciprocal discrepancy is $2.58\times10^{-6}$, and its Hessian antisymmetry residual is $6.71\times10^{-9}$. The two constant mass endpoints are solved again relative to the finer reference mass. Their surface gravity shifts differ from the standard results by at most $1.08\times10^{-6}$ parts per million, compared with the shifts of order $15$ parts per million at $\mu=0.30M_0$. Across the retained configurations, the largest absolute relative rest mass and Komar mass discrepancies are $6.45\times10^{-11}$ and $1.20\times10^{-12}$, respectively. The corresponding orbital energy integral discrepancy is $6.45\times10^{-11}$.

A third calculation uses $(3601,48,64,30,192)$ and a reconstruction tolerance of $5\times10^{-13}$. Its undeformed reference is initialized from a vacuum metric, and both endpoints are reconstructed relative to that reference. The largest endpoint mass residual is $5.33\times10^{-13}M_0$, and the largest difference from the standard surface gravity shifts is $2.42\times10^{-7}$ parts per million. The local stress balance residual in Eq.~\eqref{eq:local_stress_balance} is at most $4.66\times10^{-9}$ for these reconstructed configurations. The mass constraint residual controls equality within a given discretization; the independent refinements assess the sensitivity of the measured horizon response.

An independent limiting check follows from the dilute orbital kernel of Section~\ref{sec:benchmark}. With $\langle X\rangle_0=\int h_0X\,dJ_r\,dL$, define
\begin{equation}
 \widehat{\mathcal A}_i=\lim_{\mu\to0}\frac{\mathcal A_i}{\mu}
 =\langle Q_iE_0\rangle_0,\qquad
 \widehat{\mathcal R}_i=\lim_{\mu\to0}\frac{M_0\mathcal R_i}{\mu}
 =-\frac{M_0}{M_\bullet}\langle Q_iK\rangle_0.
 \label{eq:dilute_shape_response}
\end{equation}
Here $E_0$ and $K$ are computed in the Schwarzschild background. At $M_\bullet=M_0$, direct orbital quadrature gives
\begin{equation}
 \lim_{\mu\to0}\frac{M_0\mathcal S}{\mu}
 =\widehat{\mathcal R}_J-
 \frac{\widehat{\mathcal A}_J}{\widehat{\mathcal A}_L}
 \widehat{\mathcal R}_L=-4.83036233\times10^{-5}.
 \label{eq:dilute_constrained_response}
\end{equation}
Changing the action and orbital quadratures changes this value by $3.73\times10^{-9}$ in relative terms. A separate calculation inverts the analytic Schwarzschild action in the Darwin orbital parameters, without using the equilibrium action interpolation. It agrees with Eq.~\eqref{eq:dilute_constrained_response} to $2.60\times10^{-9}$ in relative terms. The fully coupled calculation at $\mu=0.002M_0$ gives $M_0\mathcal S/\mu=-4.8600141\times10^{-5}$, while the reference at $\mu=0.30M_0$ gives $-1.0083839\times10^{-4}$. Thus the finite occupation response includes a substantial collective field contribution to this constrained derivative.

The simultaneous Newtonian and dilute limit gives a useful contrast. The Kepler Hamiltonian written in specific actions has the standard dependence on $J_r+L$ \cite{Posti2015}:
\begin{equation}
 E_{\rm N}=1-\frac{M_\bullet^2}{2(J_r+L)^2},\qquad
 \mathcal A_i^{\rm N}=-\frac{\mu M_\bullet^2}{2}
 \left\langle\frac{Q_i}{(J_r+L)^2}\right\rangle_0,
 \qquad \mathcal R_i^{\rm N}=\frac{2\mathcal A_i^{\rm N}}{M_\bullet}.
 \label{eq:kepler_redistribution}
\end{equation}
The rest energy drops out because $\langle Q_i\rangle_0=0$. Substitution in Eq.~\eqref{eq:constant_mass_horizon_response} gives $\mathcal S^{\rm N}=0$ at this order, for any pair of admissible redistribution directions with $\mathcal A_L^{\rm N}\ne0$. The nonzero response in Eq.~\eqref{eq:dilute_constrained_response} survives in the relativistic dilute limit, and self gravity modifies it at finite $\mu$. The Schwarzschild benchmark isolates the relativistic dilute contribution, and the coupled equilibria quantify its modification by the gravity of the population.

\subsection{Orbital scale dependence and the first relativistic correction}
\label{sec:orbital_scale}

The Kepler cancellation can be expanded systematically. Define $I=J_r+L$ and keep $L/I$ bounded away from zero. The Schwarzschild energy expressed in specific actions, through the first relativistic correction, is the test particle limit of the Delaunay Hamiltonian \cite{Damour2000}:
\begin{equation}
 E_0=1-\frac{M_\bullet^2}{2I^2}
 +M_\bullet^4\left(\frac{15}{8I^4}-\frac{3}{I^3L}\right)
 +\mathcal O\!\left(\frac{M_\bullet^6}{I^6}\right).
 \label{eq:action_energy_relativistic}
\end{equation}
The remainder refers to a uniform expansion at fixed action shape. Write the Newtonian and first relativistic contributions to the occupation gradient as
\begin{equation}
 \mathcal A_i^{(0)}=-\frac{\mu M_\bullet^2}{2}
 \left\langle\frac{Q_i}{I^2}\right\rangle_0,\qquad
 \mathcal A_i^{(1)}=\mu M_\bullet^4
 \left\langle Q_i\left(\frac{15}{8I^4}-\frac{3}{I^3L}\right)\right\rangle_0.
 \label{eq:relativistic_gradient_parts}
\end{equation}
At dilute order, differentiation at fixed dimensional actions gives $\mathcal R_i=(2\mathcal A_i^{(0)}+4\mathcal A_i^{(1)})/M_\bullet$ to the retained accuracy. Expanding the ratio of gradients in Eq.~\eqref{eq:constant_mass_horizon_response} therefore yields the first nonzero constrained contribution,
\begin{equation}
 \mathcal S_{\rm 1PN}=\frac{2}{M_\bullet}
 \left(\mathcal A_J^{(1)}
 -\frac{\mathcal A_J^{(0)}}{\mathcal A_L^{(0)}}\mathcal A_L^{(1)}\right).
 \label{eq:relativistic_constrained_leading}
\end{equation}
The two orders carry different powers of the central mass, so the energy constraint does not remove both of their horizon responses. Equation~\eqref{eq:relativistic_constrained_leading} applies when $\mathcal A_L^{(0)}\ne0$ and the action support remains in the weak field regime.

\begin{figure}[!t]
\centering
\includegraphics[width=0.97\linewidth]{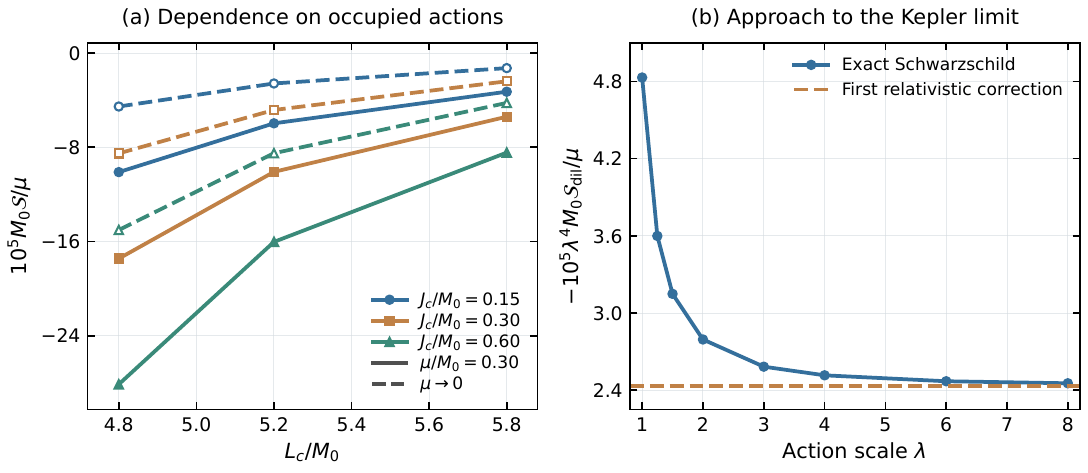}
\caption{Orbital dependence of the constrained horizon response at $M_\bullet=M_0$. (a) Solid curves join coupled calculations at $\mu/M_0=0.30$; dashed curves join independent dilute Schwarzschild calculations at the same nine action shapes. Colours and symbols identify $J_c$, with $\Delta_J=2J_c/3$ and $\Delta_L=0.30M_0$. Every point evaluates the tangent satisfying all three mass constraints at its own undeformed reference. (b) The exact dilute response for a common scaling $\lambda$ of all reference action centres and widths. Multiplication by $-\lambda^4$ exposes its approach to the positive magnitude of the leading coefficient in Eq.~\eqref{eq:relativistic_constrained_leading}. The horizontal dashed line is that coefficient. }
\label{fig:orbital_scale}
\end{figure}

For the reference shape at $M_\bullet=M_0$, Eq.~\eqref{eq:relativistic_constrained_leading} gives $M_0\mathcal S_{\rm 1PN}/\mu=-2.43469391\times10^{-5}$. This leading term is an asymptotic explanation; the reference population still requires the exact relativistic orbital calculation. To test the expansion, multiply all action centres and widths in Eq.~\eqref{eq:reference_family} by a common factor $\lambda$, while keeping the dimensionless modes and $M_\bullet=M_0$ fixed. The leading constrained response scales as $\lambda^{-4}$. Figure~\ref{fig:orbital_scale}(b) compares eight exact Schwarzschild quadratures with that prediction. At $\lambda=8$, the fractional difference between the leading term and the exact dilute response is $0.815$ per cent; changing the action and orbital quadratures changes the exact result by $1.77\times10^{-12}$ in relative terms.

The finite population dependence is examined separately in Fig.~\ref{fig:orbital_scale}(a). We reconstruct nine reference populations at $\mu=0.30M_0$, using $J_c/M_0=0.15,0.30,0.60$ and $L_c/M_0=4.8,5.2,5.8$, with $\Delta_J=2J_c/3$ and $\Delta_L=0.30M_0$. Each point uses six additional neighboring equilibria for the two occupation responses and the central mass derivative. The largest relative mass gradient and reciprocal discrepancies over this grid are $5.54\times10^{-5}$ and $3.78\times10^{-6}$, respectively. The constrained response is negative at all nine points and lies between $-2.81\times10^{-4}$ and $-3.28\times10^{-5}$ in $M_0\mathcal S/\mu$. Its magnitude decreases as the mean angular momentum increases at each of the sampled radial action centres. Exact dilute responses at the same action shapes show that the population field enhances the magnitude over this grid. These comparisons resolve the parameter dependence of the chosen family without extrapolating its sign to arbitrary occupations.

\section{Discussion}
\label{sec:discussion}

The constrained families in Section~\ref{sec:constant_mass_curve} isolate information that is not fixed by global mass data. Matter outside a spherical horizon can alter its asymptotic normalization even when the horizon and ADM masses are prescribed, as illustrated by the thin shell construction of Ref.~\cite{Visser1992}. The present calculation imposes the additional particle rest mass constraint and realizes the exterior matter as a distributed, self consistent kinetic source. Its radial and tangential stresses are moments of the same occupation function that determines the orbital energies, and every point on the redistribution curve is a solution of the coupled equilibrium problem. The result is therefore not the statement that exterior matter can redshift a horizon, but that the three masses $(M_\bullet,M_{\rm rest},M_\infty)$ do not uniquely determine the inner Killing normalization within this class of Einstein--Vlasov equilibria.

There is no conflict between fixed horizon area and varying surface gravity in Eq.~\eqref{eq:constant_area_surface_gravity}. The common horizon mass fixes the Schwarzschild curvature in the inner vacuum interval, but it does not fix the rate of the local Killing time relative to infinity. Within each constant mass family, the outer vacuum metric has the same ADM mass, while the distribution through the matter region changes. Figure~\ref{fig:constant_mass}(c,d) resolves this distinction: the enclosed mass differences vanish outside the matter, whereas the lapse ratio differences approach nonzero constants in the inner vacuum interval. Surface gravity refers throughout to the Killing field normalized at infinity. The calculated change is consequently a change of this global normalization rather than a change of the local vacuum curvature adjacent to the horizon.

The derivative identities distinguish two different comparisons. An adiabatic change of the central mass preserves the occupation of each action cell. Redistribution at fixed rest mass changes those occupations and compares neighbouring stationary populations. Its mass gradient is the orbital energy contrast in Eq.~\eqref{eq:shape_energy_gradient}, including the induced field response. On the locally regular branch used here, Eq.~\eqref{eq:reciprocity_weak} states the corresponding mixed derivative symmetry as an integrated identity for smooth occupation directions. The finite dimensional mode tests in Section~\ref{sec:selfgravity} are direct projections of that identity and do not require a stronger pointwise differentiability assumption.

The limiting calculation clarifies which part of the response requires relativistic orbital structure. At leading order in a dilute Kepler population, the horizon response gradients are proportional to the energy gradients for every admissible occupation mode, so the projection that fixes the ADM mass cancels. Exact Schwarzschild orbits remove that proportionality and give the nonzero value in Eq.~\eqref{eq:dilute_constrained_response}. Equations~\eqref{eq:relativistic_gradient_parts} and \eqref{eq:relativistic_constrained_leading} identify the leading mismatch and its fourth inverse power dependence on a common action scale. The finite population calculation is separate: its departure from the dilute value contains the response of the self gravitational field. The relativistic dilute contribution and the finite mass modification are therefore identified by distinct calculations rather than conflated into a single environmental correction.

The numerical shifts in Table~\ref{tab:constant_mass} and the orbital comparisons in Fig.~\ref{fig:orbital_scale} characterize the specified compact action families. Their magnitudes are resolved relative to the reported mass constraint residuals, derivative discrepancies and independent refinements. The calculation establishes the existence and size of the constrained response for these regular Einstein--Vlasov environments; it does not rely on interpreting the chosen reference occupation as a fit to a particular galactic density profile. For a specified astrophysical population, the same formalism can instead be applied to an occupation inferred from that system. Within the equilibrium problem considered here, the central conclusion is already fixed by the source consistent construction: orbital occupation contains horizon normalization information not determined by the three global mass constraints.

\section{Conclusions}
\label{sec:conclusions}

We have formulated the response of static spherical black holes with collisionless matter in terms of the occupation of regular bound orbital actions. Linearizing the spherical mass constraint gives a kinetic variation law in which the orbital energy measured at infinity weights an occupation change and the inner lapse ratio multiplies the horizon mass variation. On a locally regular equilibrium branch, commuting the central mass and occupation variations gives an integrated reciprocity relation connecting the mass derivative of orbital energy to the directional response of the horizon normalization.

The coupled calculations verify these relations for compact finite mass populations. Changes of the mean radial action, mean angular momentum and radial action variance provide independent occupation directions; their orbital energy integrals reproduce the corresponding ADM mass derivatives. Local stress balance, conserved mass integrals, action reconstruction, derivative comparisons, Hessian symmetry and independent resolution changes provide separate consistency tests.

Using these occupation gradients, we trace families with fixed horizon mass, particle rest mass and ADM mass. The horizon area is constant, while the surface gravity changes through the inner lapse ratio. For the reference rest mass ratio $0.30$, the endpoint changes are approximately $+15.37$ and $-15.33$ parts per million relative to the undeformed equilibrium. The same constrained response cancels at simultaneous dilute and Newtonian order, is nonzero for exact Schwarzschild orbits, and is modified when the population field is included. The first relativistic correction accounts for the leading orbital scale dependence. Thus, within the regular Einstein--Vlasov families constructed here, the three global mass constraints do not uniquely determine the asymptotic normalization of the horizon Killing field.

\end{document}